\documentclass[11pt,a4paper]{article}
\pdfoutput=1

\usepackage{jcappub}
\usepackage[latin1]{inputenc}
\usepackage{graphicx}
\usepackage{epsfig}
\usepackage{subfigure}
\usepackage{color}
\usepackage{comment}
\usepackage{slashed}
\usepackage{soul}
\usepackage{tabularx}

\newcommand{\be}{\begin{equation}} 
\newcommand{\ee}{\end{equation}}
\newcommand{\bea}{\begin{equation}\begin{aligned}} 
\newcommand{\eea}{\end{aligned}\end{equation}}

\def\lsim{\mathrel{\raise.3ex\hbox{$<$\kern-.75em\lower1ex\hbox{$\sim$}}}}
\def\gsim{\mathrel{\raise.3ex\hbox{$>$\kern-.75em\lower1ex\hbox{$\sim$}}}}

\usepackage{array}
\newcolumntype{C}[1]{>{\centering\let\newline\\\arraybackslash\hspace{0pt}}m{#1}}

\newcommand{\GeV}{{\rm GeV}}

\newcommand{\eV}{{\rm eV}}

\newcommand{\mpl}{M_{\rm P}}

\newcommand{\eg}{\emph{e.g.\,}}
\newcommand{\ie}{\emph{i.e.\,}}

\newcommand{\td}{{\rm d}}
\newcommand{\pd}{\partial}
\newcommand{\Hinf}{H_{\rm inf}}
\newcommand{\Peq}{P_{\rm eq}}

\begin{document}

\title{Primordial black holes from \\ spectator field bubbles}

\author[a,b]{David Navidad Maeso,}
\author[a]{Luca Marzola,}
\author[a]{Martti Raidal,}
\author[c]{Ville Vaskonen}
\author[]{and}
\author[a]{Hardi Veerm\"ae}

\affiliation[a]{NICPB, R\"avala pst. 10, 10143 Tallinn, Estonia}
\affiliation[b]{Tallinn University, School of Natural Sciences and Health, Narva mnt. 29, 10120 Tallinn, Estonia
}
\affiliation[c]{Institut de Fisica d'Altes Energies, The Barcelona Institute of Science and Technology, Campus UAB, 08193 Bellaterra, Barcelona, Spain}

\emailAdd{david.navidad.maeso@kbfi.ee} 
\emailAdd{luca.marzola@cern.ch}
\emailAdd{martti.raidal@cern.ch}
\emailAdd{hardi.veermae@cern.ch}
\emailAdd{vvaskonen@ifae.es}

\abstract{We study the evolution of light spectator fields in an asymmetric polynomial potential. During inflation, stochastic fluctuations displace the spectator field from the global minimum of its potential, populating the false vacuum state and thereby allowing for the formation of false vacuum bubbles. By using a lattice simulation, we show that these bubbles begin to contract once they re-enter the horizon and, if sufficiently large, collapse into black holes. This process generally results in the formation of primordial black holes, which, due to the specific shape of their mass function, are constrained to yield at most 1\% of the total dark matter abundance. However, the resulting population can source gravitational wave signals observable at the LIGO-Virgo experiments, provide seeds for supermassive black holes or cause a transient matter-dominated phase in the early Universe.
}

\maketitle

\section{Introduction}
\label{sec:intro}

In the simplest scenarios, the Universe underwent an epoch of exponential expansion driven by the dynamics of a single scalar field: the inflaton. The primordial perturbation spectrum observed in the cosmic microwave background is thus commonly ascribed solely to the properties of this particle. However, it is still possible that additional degrees of freedom leave their imprints on the dynamics consequent to the inflationary expansion even when they do not drive the latter. These fields -- collectively known as spectator fields -- are at the basis of well-known frameworks such as that of the curvaton~\cite{Lyth:2001nq,Moroi:2001ct,Enqvist:2001zp,Linde:1996gt,PhysRevD.42.313} and modulated reheating scenarios~\cite{Kofman:2003nx,Dvali:2003em,Ichikawa:2008ne}, which can source the observed perturbations in place of the inflaton. Given the inferred limit on the scale of inflation, in recent years a considerable effort was also dedicated to the investigation of the Higgs boson as a concrete realization of spectator field~\cite{Choi:2012cp, DeSimone:2012gq, DeSimone:2012qr, Cai:2013caa, Freese:2017ace, Espinosa:2017sgp, Espinosa:2018eve, Lu:2019tjj, Litsa:2020mvj, Litsa:2020rsm, Karam:2020skk, Karam:2021qgn}.    

Broadly speaking, spectator fields are scalar fields characterised by energy densities well below that of the inflaton during the exponential expansion epoch and, consequently, have only a negligible bearing on the latter. The evolution of spectator fields can be studied within a quantum field theory built on the approximately de Sitter background set by the inflaton~\cite{Chernikov:1968zm,Dowker:1975tf,Bunch:1978yq,Birrell:1982ix,Linde:1982uu,Allen:1985ux,Allen:1987tz}, however the full characterization of their dynamics often requires approaches beyond the usual perturbative methods~\cite{Hu:1986cv,Boyanovsky:2005sh,Serreau:2011fu,Herranen:2013raa,Gautier:2013aoa,Gautier:2015pca,Tokuda:2017fdh,Arai:2011dd,Guilleux:2015pma,Prokopec:2017vxx,Moreau:2018ena,Moreau:2018lmz,LopezNacir:2019ord}. 

Alternatively, the evolution of these fields can be modelled by using a stochastic method which returns a probability distribution of the spectator field values~\cite{Starobinsky:1986fx,Starobinsky:1994bd}. The approach consists in integrating out the quantum evolution of sub-horizon scalar field modes, thereby obtaining a stochastic term that is used as a source for the approximately classical evolution of the remaining super-horizon modes. The resulting Langevin equation shows that spectator fields exhibit stochastic fluctuations of their field values, and the stationary solution of the associated Fokker-Planck (FP) equation then gives the corresponding equilibrium distribution. For practical purposes, it is often convenient to rewrite the obtained FP equation as an eigensystem problem, in a way that arbitrary correlation functions involving spectator field values, as well as the corresponding power spectra, can be obtained via a spectral expansion~\cite{Markkanen:2019kpv,Markkanen:2020bfc}.

The stochastic formalism shows that during the inflationary expansion, a spectator field fragments into a set of coherent domains, each characterised by a field value drawn by the one-point probability distribution. When the effective mass of a spectator field is below the Hubble parameter set by the inflation scale, the effective friction prevents the evolution of the spectator field and such fragmentation is therefore maintained until after the end of inflation.

In the present paper, we use this effect to propose a new mechanism of primordial black hole (PBH) formation, proceeding from the collapse of false vacuum bubbles (FVBs) of a spectator field. In more detail, considering a real scalar field with an asymmetric double-well potential, we show that the stochastic evolution of the spectator field displaces it from its true vacuum state and results in the formation of domains where the field is trapped into a false vacuum. Through a lattice simulation, we then compute the evolution of these domains once they re-enter the horizon after inflation assuming either a radiation-dominated (RD) or a matter-dominated (MD) universe. The results show that FVBs collapse and, if sufficiently large, form black holes. 

PBHs have a potential impact on a large variety of physical phenomena. For instance, stellar mass PBHs are potential progenitors~\cite{Bird:2016dcv,Sasaki:2016jop,Clesse:2016vqa,Raidal:2017mfl,Hutsi:2020sol,DeLuca:2021wjr} of the BH-BH collisions observed by the LIGO-Virgo collaboration~\cite{LIGOScientific:2018mvr,LIGOScientific:2020ibl} and heavier PBHs can act as seeds for supermassive BHs~\cite{Bean:2002kx,Kawasaki:2012kn,Clesse:2015wea,Bernal:2017nec}. PBHs in the asteroid mass window $10^{17}-10^{23}\, \rm g$ also constitute a viable dark matter (DM) candidate~\cite{Carr:2020gox}. Importantly, even BHs lighter than $10^{9}\, \rm g$, evaporating before big-bang nucleosynthesis (BBN)~\cite{Carr:2009jm}, can still substantially affect DM phenomenology~\cite{Fujita:2014hha,Allahverdi:2017sks,Lennon:2017tqq,Hooper:2019gtx,Masina:2020xhk,Baldes:2020nuv,Khlopov:2004tn,Gondolo:2020uqv,Morrison:2018xla,Bernal:2020bjf,Bernal:2020kse,Bernal:2020ili}.

The possible relation between PBH formation and spectator field dynamics has been previously studied in scenarios where the fluctuations of these fields induce large density perturbations which gravitationally collapse into BHs after re-entry~\cite{Espinosa:2017sgp}. Instead, in the case of FVBs, the collapse is mainly driven by the scalar field dynamics. Our study then adds the collapse of FVBs generated from the stochastic evolution of spectator fields to the list of possible PBH formation mechanisms (for a review see \eg~Refs.~\cite{Khlopov:2008qy,Carr:2020gox}), delineating a novel framework complementary to those based on the collapse of vacuum bubbles nucleated during inflation~\cite{Garriga:2015fdk,Deng:2017uwc,Deng:2020mds,Kusenko:2020pcg} or post-inflationary scalar field fragmentation~\cite{Cotner:2018vug,Cotner:2016cvr,Cotner:2019ykd}.

The paper is organised as follows: in Section~\ref{sec:spect}, we present the stochastic formalism for spectator field fluctuations and estimate the distribution of FVBs created during inflation. Section~\ref{sec:FVBs} treats the evolution of FVBs and formulates the conditions for FVB to collapse into BHs. In section~\ref{sec:PBHs}, instead, we estimate the properties of the resulting PBH population. We conclude with section~\ref{sec:concl} and present technical details concerning the adopted formalism in the appendix~\ref{app:FP}.

\section{Spectator field fluctuations}
\label{sec:spect}

The super-horizon scale evolution of a spectator field in a de Sitter background can be modelled as a stochastic process after quantum fluctuations at smaller scales have been integrated out. The dynamics of long wavelength modes then follows the Langevin equation~\cite{Starobinsky:1986fx,Starobinsky:1994bd}
\be\label{eq:langevin}
    \dot{\phi} = -\frac{1}{3\Hinf} V'(\phi) + \xi,
\ee
where $V(\phi)$ is the spectator field potential, a prime denotes differentiation with respect to $\phi$, $\Hinf$ is the Hubble parameter during inflation, and $\xi$ is white noise produced by the quantum fluctuations. The white noise is characterized by the correlator $\langle \xi(t,\vec x)\xi(t',\vec x')\rangle \approx \Hinf^3/(2\pi)^2\delta(t-t')\theta(1 - \epsilon a H (\vec x - \vec x'))$, where $\epsilon$ is a small (but not infinitesimal) parameter, and thus is uncorrelated at proper distances larger than $(\epsilon H)^{-1}$. The overdamped Langevin equation~\eqref{eq:langevin} does not couple the field at different spatial locations. Therefore, in order to derive the spatial correlations, it must be assumed that the field is strongly correlated at physical distances $\lesssim(\epsilon H)^{-1}$.

\begin{figure}
\centering
\includegraphics[width=0.6\textwidth]{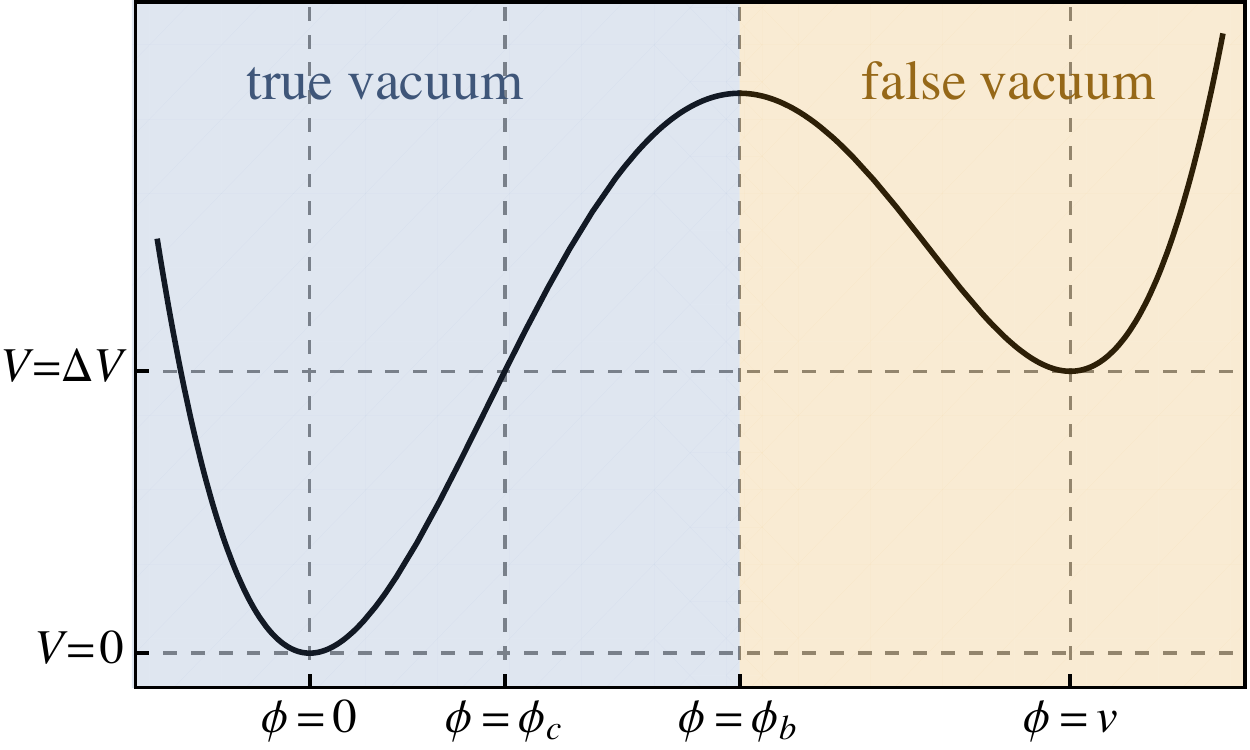}
\caption{The asymmetric double-well scalar potential of the spectator field.}
\label{fig:pot}
\end{figure}

The spectator field we consider is a real scalar field with a general quadratic potential parametrized as
\be \label{eq:pot}
    V(\phi) = \Delta V \left[(\kappa+6)\left(\frac{\phi}{v}\right)^2 - 2(\kappa+4)\left(\frac{\phi}{v}\right)^3 + (\kappa+3)\left(\frac{\phi}{v}\right)^4 \right] \,,
\ee
where $\Delta V, v, \kappa > 0$. As illustrated in Fig.~\ref{fig:pot}, the potential has a global minimum at $\phi = 0$ with $V(0) = 0$ and a local minimum at $\phi = v$, with $V(v) = \Delta V$. The height of the barrier that separates the minima is
\be\label{eq:Vmax}
    V(\phi_b) = \frac{(\kappa+2)(\kappa+6)^3}{16 (\kappa+3)^3} \,\Delta V,
    \qquad \mbox{with} \qquad 
    \phi_b = \frac{\kappa+6}{2 (\kappa+3)} \, v \,.
\ee 
The parameter $\kappa$ quantifies the size of the barrier between the two vacuum states: for $\kappa=0$ the barrier vanishes and the potential presents an inflection point at $\phi = v$, while for $\kappa \gg 1$ the potential is nearly symmetric with respect to $\phi_b \approx v/2$.

For the stochastic fragmentation of the field to be maintained until after the end of inflation, the Hubble friction must prevent the evolution of the long-range modes, that is:
\be\label{eq:m<H}
    m^2_{\rm eff}(\phi) \equiv V''(\phi) \lesssim \Hinf^2.
\ee

\subsection{Evolution of the probability distribution}

The statistical properties of the spectator field within a Hubble patch are determined by the FP equation corresponding to the Langevin equation~\eqref{eq:langevin},
\be\label{eq:FP}
    \partial_{t} P 
    = \mathcal{L}_{\rm FP} P
    \equiv \frac{1}{3 \Hinf}\partial_{\phi}\left( V' P \right) + \frac{\Hinf^3}{8\pi^2}\partial_{\phi}^2P\,,
\ee
where $\mathcal{L}_{\rm FP}$ denotes the FP operator. The FP equation admits a unique equilibrium distribution
\be\label{eq:P_eq}
    \Peq(\phi) \propto \exp[-8\pi^2 V(\phi)/(3\Hinf^4)]
\ee 
provided that the scalar potential grows at large field values. The spatial profile of a spectator field can then be derived from its temporal evolution by using de Sitter invariance, which maps time intervals to equivalent spatial intervals~\cite{Markkanen:2018gcw, Markkanen:2019kpv, Markkanen:2020bfc, Karam:2020skk}.

Given an initial distribution $P_{\rm i}(\phi',t_0)$, the FP equation~\eqref{eq:FP} can be integrated to give
\be
    P(\phi,t) = \int \td \phi' P\left(\phi,t| \phi',t_0\right)P_{\rm i}(\phi',t_0),
\ee
as is the case for any Markov process. Solving the eigenvalue problem of the FP operator, $\mathcal{L}_{\rm FP}  P_n = \Lambda_n P_n$, allows for a spectral expansion of the transition probability~\cite{Risken},
\be\label{eq:trans_Pt}
    P(\phi,t|\phi',t') 
    = \frac{1}{\Peq(\phi')}  \sum_{n\geq 0} P_n(\phi) P_n(\phi') e^{-\Lambda_n (t-t')}. 
\ee 
The eigenfunctions $P_n(\phi)$ are complete, i.e. $P(\phi,t|\phi',t) = \delta(\phi - \phi')$, and orthogonal with respect to the measure $\td \phi/\Peq$.\footnote{
The FP operator satisfies
\be
    \Peq^{-1} \left( P_1 \mathcal{L}_{\rm FP} P_2 - P_2 \mathcal{L}_{\rm FP} P_1\right) = \partial_{\phi}\left[\Peq^{-1} \frac{H^3}{8\pi^2} \left(P_1 \partial_{\phi} P_2 - P_2 \partial_{\phi} P_1\right)\right]
\ee
and is thus Hermitean with respect to the scalar product $(P_1,P_2) \equiv \int \frac{\td \phi}{\Peq} P_1 P_2$ when $P_i$ (or $\partial_{\phi} P_i$) vanish at the boundary of the domain of $P_i$. As a result, the eigenfunctions of $\mathcal{L}_{\rm FP}$ form an orthogonal basis. When $\partial_{\phi} P_i$ does not vanish at the boundary, as is the case with absorbing boundary conditions, the probability can flow out of the domain and ceases to be conserved.
}
We choose the normalization so that $\int \td \phi P_n(\phi) P_m(\phi)/\Peq(\phi) = \delta_{nm}$ and label the states in order of increasing eigenvalues. In the domain $\phi \in \mathbb{R}$, the lowest eigenvalue vanishes and $P_0 = \Peq$. To compute the eigenvalues, we solve the equivalent Schr\"odinger equation given in appendix~\ref{app:FP}.

Depending on the processes we aim to describe, the FP equation~\eqref{eq:FP} can be solved with different boundary conditions. We consider three possibilities:
\begin{itemize}
    \item Non-absorbing boundary conditions: $P(\phi,t) \to 0$ when $\phi \to \pm \infty$.
    \item Absorbing boundary conditions in the false vacuum region: $P_F(\phi,t) = 0$, $\phi \geq \phi_c$ and $P_F(\phi,t) \to 0$ when $\phi \to \infty$. We denote quantities computed using these boundary conditions by the subscript $F$.
    \item Absorbing boundary conditions in the true vacuum region: $P_T(\phi,t) = 0$, $\phi \leq \phi_c$ and $P_T(\phi,t) \to 0$ when $\phi \to -\infty$. In this case, all quantities are denoted by the subscript $T$.
\end{itemize}
The absorbing boundary conditions can be used to describe the first transition from the true to the false vacuum (or vice versa). In the absorbing cases, there is a probability flux across the boundary, so the total probability is not conserved. In particular, the lowest eigenvalue will be strictly positive~\cite{Risken}. The total probability corresponds to the probability that the FVB has survived. Although the eigen-decomposition of the transition probability~\eqref{eq:trans_Pt} can be always be achieved in the considered cases, the eigenfunctions $P_{n}$ must satisfy the imposed boundary conditions and the eigenvalue spectra will generally differ.

In the stochastic formalism, the sub-horizon field is assumed to be almost completely correlated. Then, every Hubble patch will contain an homogeneous field with a random value drawn by the field value distribution. As the Universe expands, the comoving size of the horizon shrinks and the field in the initial Hubble patch begins to fluctuate. The equilibration time, understood as the time required to completely wash out the initial value of the field, will generally differ across different domains.

Let us discuss the largest patch currently observable, which exited the horizon at about 60 $e$-folds before the end of inflation. Because the value of the homogeneous field at the time of horizon crossing is not known, we assume $\phi = 0$ as indicated by the field value equilibrium distribution. Thus $P_{\rm i} = \delta(\phi)$ and
\be\label{eq:P(N)}
    P(\phi,N) 
    = P\left(\phi,N| 0,N_i\right)
    = \Peq(\phi) + \sum_{n\geq 1} \frac{P_n(0)}{\Peq(0)} P_n(\phi) e^{-\frac{\Lambda_n}{H} (N-N_i)}\, ,
\ee
where $N \approx H t$ is the number of $e$-folds, the currently observable patch exited the horizon at $N=N_i$ and we used non-absorbing boundary conditions so that $P_{0} = \Peq$. For practical purposes, the series can be truncated at the lowest $n$ such that $N-N_i \ll H/\Lambda_n$ and the observable Hubble patch is thus in equilibrium when $N-N_i \gg H/\Lambda_1$. If the barrier that separates the two vacua is large, then the lowest non-vanishing eigenvalue $\Lambda_1$ is necessarily suppressed with respect to the next eigenvalues. In fact, $\Lambda_1$ controls the transitions (and consequent equilibration) between true and false vacuum, whereas the higher eigenvalues describe mainly the fluctuations around the two minima.

In the following, we consider scenarios in which the two minima are separated by a sizable barrier such that
\be\label{eq:cond_V}
    V(\phi_b) \gtrsim \frac{3}{4\pi^2}\Hinf^4, 
    \qquad
    \kappa \gg 1.
\ee
The first condition implies that fluctuations over the barrier are rare. The second condition, instead, guarantees that the value of potential evaluated at the top of the barrier is considerably larger than the energy of false vacuum, \ie that $V(\phi_b) > \Delta V$. If both conditions are met, then, at the equilibrium~\eqref{eq:P_eq}, the field will mostly occupy regions close to the minima of its potential~\eqref{eq:pot}. Moreover, transitions between the two vacua are unlikely and therefore it is safe to assume that equilibration is only achieved locally, around the two minima. For $\kappa \gg 1$, the requirement that the field is light~\eqref{eq:m<H} and the condition~\eqref{eq:cond_V} yield
\be\label{eq:cond_V2}
    \frac{12}{\pi^2} \lesssim \kappa \frac{\Delta V}{\Hinf^4} \lesssim \frac{v^2}{2\Hinf^2} 
\ee
so $v$ is necessarily larger than the scale of inflation.

In the following, we will take a closer look at how the field occupies the false and true vacuum regions under the assumptions we have discussed.

\subsection{Fluctuations into the false vacuum}

As the spectator field is initially in its true vacuum, each Hubble patch in which the field fluctuates to the false vacuum region generates a potential FVB\footnote{This process is the stochastic equivalent of true vacuum decay in de Sitter space~\cite{Lee:1987qc}. Tunneling has been previously considered, \eg, in the context of stochastic inflation \cite{Starobinsky:1994bd,Noorbala:2018zlv}.}. The comoving size of the bubble then roughly matches that of the horizon at the time of the transition
\be\label{eq:R0_a_f}
    R_0 = \frac{1}{a_f H(a_f)},
\ee
where $a_f$ denotes the scale factor at the time the FVB was formed. In order to derive the FVB size distribution, we follow the argument presented in Ref.~\cite{Deng:2017uwc} for the nucleation and expansions of TVBs, accounting for the fact that in our case the entire Hubble patch fluctuates into the false vacuum region and the dynamics subsequently freezes until re-entry.

The fraction of true vacuum patches transitioning to the false vacuum in a time interval $\Delta t$ is $1-P_{T}(\phi < \phi_b,t+\Delta t| \phi_0, t)$. Since we are interested in the first crossing, the transition probability is promptly computed by using absorbing boundary conditions at $\phi = \phi_b$ and therefore we define the probability of FVB formation within a Hubble time as
\be
    \lambda_{T \to F} \equiv -\Hinf^{-1}\lim_{t' \to t}\int^{\phi_b}_{-\infty} \td \phi' \frac{\td }{\td t'}P_T(\phi',t'| \phi_0, t).
\ee
The number of FVBs in the volume $\td V$ is proportional to the rate $\lambda_{T \to F}$ and to the number of Hubble patches, which is given by $(3/4\pi) \Hinf^3\td V$. At formation, $\td V = a_f^3 \td^3 x = \td^3 x (R_0 \Hinf)^{-3}$, while the time of formation and the FVB size are related through $\td t = \td R_0 (R_0 \Hinf)^{-1}$. Consequently, the comoving number density of FVBs with a comoving size in the range $(R_0,R_0+\td R)$ is 
\be\label{eq:nBnoneq}
    \td n_{\rm B}(R_0) = \lambda_{T \to F} \frac{\td R_0}{4\pi R_0^{4}/3}.
\ee
We remark that the time evolution of the transition probability $\lambda_{T \to F}$ can generate deviations from the $R_0^{-4}$ scaling as the bubble size $R_0$ is directly related to the time via Eq.~\eqref{eq:R0_a_f}.

\begin{figure}
\centering
\includegraphics[width=0.7\textwidth]{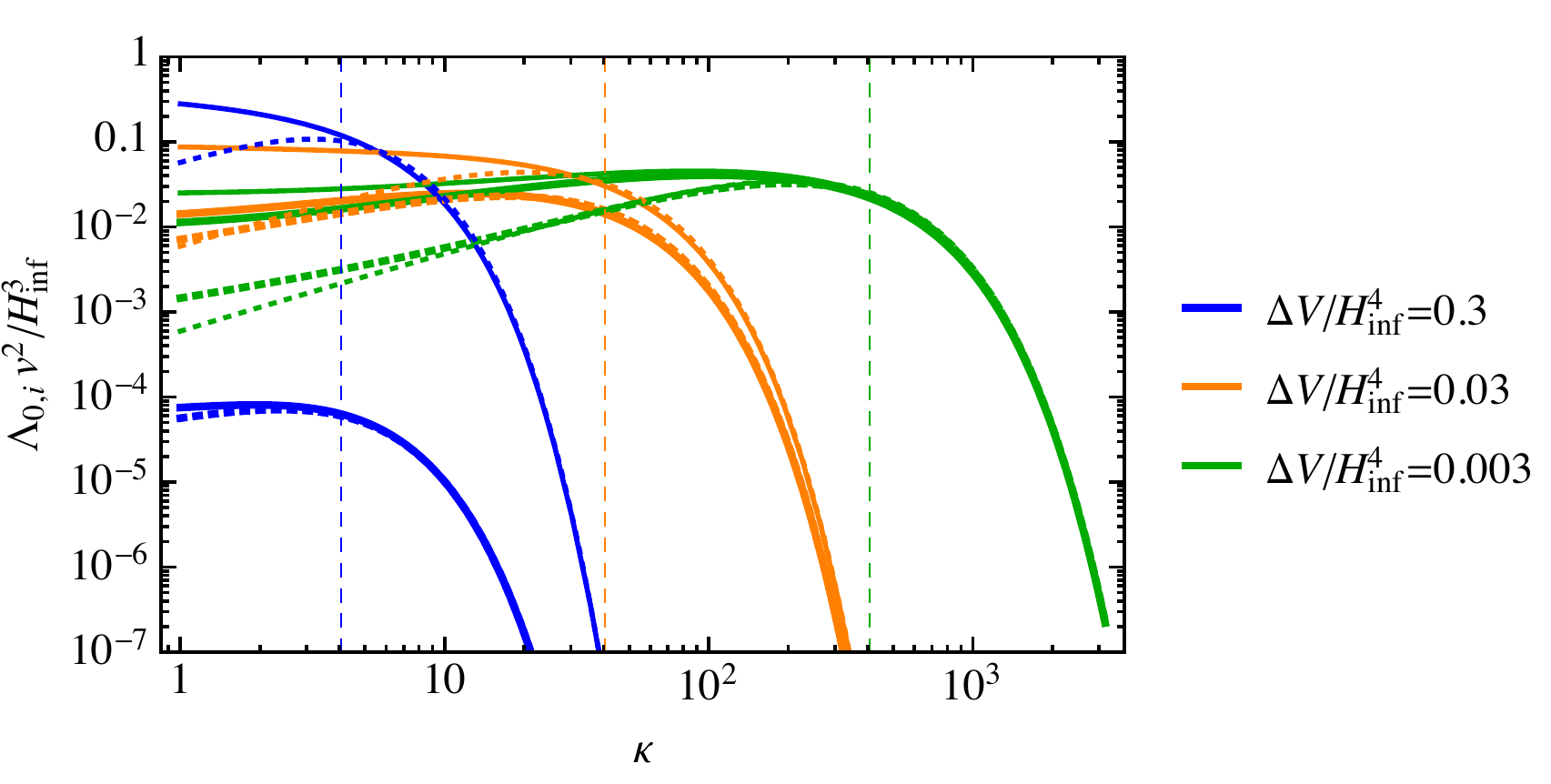}
\caption{The 0-th eigenvalue of the FP operator with absorbing boundary conditions in the true vacuum region (thick lines) and in the false vacuum region (thin lines). The solid curves are obtained through a numerical computation while the dashed curves use the approximations in Eqs.~\eqref{eq:Lambda0TV} and~\eqref{eq:Lambda0FV}. For each of the analyzed cases, the condition~\eqref{eq:cond_V} is fulfilled in the region on the right of the corresponding vertical dashed line.}
\label{fig:Lambda0}
\end{figure}

After a sufficient number of $e$-folds, the transition probability in Eq.~\eqref{eq:trans_Pt} is determined only by the lowest eigenvalues, hence at the first order
\be
    \lambda_{T \to F} \approx \Lambda_{0,T}/H
\ee
where $\Lambda_{0,T}$ is the leading eigenvalue. In appendix~\ref{app:FP}, we show that $\Lambda_{0,T}$ can be approximated as
\be \label{eq:Lambda0TV}
    \Lambda_{0,T} \approx \frac{\sqrt{-V''(0) V''(v_b)}}{3\pi \Hinf} \exp\left[ -\frac{8\pi^2}{3 \Hinf^4} V(\phi_b) \right] \,,
\ee
and in Fig.~\ref{fig:Lambda0} we compare the above relation (dashed curves) with the numerical computation of $\Lambda_{0,T}$ (solid lines) for different values of the parameters in the potential.

\subsection{Fluctuations into the true vacuum}

The interior of a FVB may fluctuate back to the true vacuum and the resulting TVBs, if sufficiently large, would then expand after re-entering the horizon. As such TVBs may destroy the surrounding FVB before it collapses to a black hole, we disregard FVBs that contain such TVBs when estimating the PBH population. Nevertheless, similarly to PBH production in first-order phase transitions~\cite{Hawking:1982ga,Kodama:1982sf}, colliding TVBs could themselves produce PBHs inside a larger FVB. By disregarding this possibility we therefore obtain a conservative estimate of the PBH abundance. 

To study the lifetime of the FVBs formed by the spectator field, we consider the time of first exit from the false vacuum region, $\phi \geq \phi_b$. To this purpose, we solve the FP equation in the region $\phi \geq \phi_b$ imposing absorbing boundary conditions at $\phi = \phi_b$, \ie~, $P_F(\phi_b) = 0$.

The false vacuum domain will grow to contain more Hubble patches as it expands. Given that each patch can fluctuate into the true vacuum independently, the probability that all sub-patches remain in the false vacuum until the time $t$ satisfies
\be\label{eq:FVB_survival}
    \Hinf^{-1} \frac{\td P_{\rm nucl, inf}}{\td t}
      = -\left(\frac{a}{a_f}\right)^3 \lambda_{F \to T} P_{\rm nucl, inf}
\ee
with $t_f$ being the time the initial patch exits the horizon, $a_f = a(t_f)$, $P_{\rm nucl, \rm inf}(t_f) = 1$, and the $a^3$ factor accounting for the expanding bubble size. The relevant transition probability within a Hubble time can be estimated as
\be \label{eq:lambda_FT}
    \lambda_{F \to T} 
    \equiv -\Hinf^{-1}\lim_{t' \to t}\int^{\infty}_{\phi_b} \td \phi'\,\td \phi\, \frac{\td }{\td t'}P_F (\phi',t'| \phi, t) P_F(\phi)
   \approx \Lambda_{0,F}/\Hinf \,,
\ee
where the integral over $\phi$ accounts for the fluctuations around the false vacuum. The last approximation in Eq.~\eqref{eq:lambda_FT} holds identically after the distribution of the field values~\eqref{eq:P(N)} has relaxed to the state corresponding to the lowest eigenvalue, \ie when $P_F(\phi) = P_{0,F}(\phi)$. We find that $\Lambda_{0,F}$ can be approximated by (see appendix~\ref{app:FP})\footnote{The rate $\lambda_{F \to T}$ can also be estimated from the principle of detailed balance: at the equilibrium $P_{{\rm eq}, F} \lambda_{F \to T}  = P_{{\rm eq}, T} \lambda_{T \to F}$. Here $P_{{\rm eq}, F} \equiv \int^{\infty}_{\phi_b} \td \phi P_{{\rm eq}}(\phi)$, $P_{{\rm eq}, T} \equiv \int^{\phi_b}_{-\infty} \td \phi P_{{\rm eq}}(\phi)$ are the equilibrium probabilities that the field is in the false or true vacuum. Thus
\be
    \lambda_{F \to T} = \lambda_{T \to F} \frac{P_{{\rm eq}, T}}{P_{{\rm eq}, F}} \approx \lambda_{T \to F} \exp\left( \frac{8\pi^2 \Delta V}{3\Hinf^4} \right).
\ee
This approach works well when the field is in near-equilibrium around the local minimum, \ie when Eq.~\eqref{eq:cond_V} is satisfied.}
\be \label{eq:Lambda0FV}
    \Lambda_{0,F} \approx \frac{\sqrt{-V''(v) V''(v_b)}}{3\pi \Hinf} \exp\left[ -\frac{8\pi^2}{3 \Hinf^4} \left(V(\phi_b)-\Delta V\right) \right] \,.
\ee
In Fig.~\ref{fig:Lambda0} we show with the solid thin curves the numerical solutions for $\Lambda_{0,F}$ and compare it to the above approximation, indicated by the dashed curves. Notice that since the condition~\eqref{eq:cond_V} requires $8\pi^2 \Delta V \lesssim 3\Hinf^4$, the transition probabilities $\lambda_{T \to F}$ and $\lambda_{F \to T}$ differ at most by an $\mathcal{O}(1)$ factor.

By integrating Eq.~\eqref{eq:FVB_survival} we obtain
\be
    P_{\rm nucl, \rm inf} (a)
    = \exp\left( - \lambda_{F \to T} \frac{a^3}{3a_f^3}  \right),
\ee
where the scale factor corresponds to a specific FVB size via Eq.~\eqref{eq:R0_a_f}. As the formalism applies during inflation, we have $a<a_{\rm inf}$ where $a_{\rm inf}$ denotes the end of inflation and $R_{\rm inf} = 1/(\Hinf a_{\rm inf})$ thus gives the size of the smallest FVBs produced. 

Depending on the scalar potential, the surface tension can cause the collapse of sufficiently small TVB nucleated inside a FVB, after the latter re-enters the horizon. This sets a critical radius $R_{*}$ and the corresponding scale factor. Defining $\bar R_* \equiv \max(R_{\rm inf},R_{*})$, we estimate that the size distribution~\eqref{eq:nBnoneq} is modified as
\be
    \td n_{\rm B}(R_0) 
    = \lambda_{T \to F} \frac{3}{4\pi}
    \frac{\td R_0}{R_0^{4}}\exp\left( -\lambda_{F \to T} \frac{R_0^3}{3{\bar R}_{*}^3} \right),
\ee
in order to account for the contamination of FVBs by contained expanding TVBs. The maximal size of FVBs that could collapse to BHs is thus approximately
\be\label{eq:R_max}
    R_{\rm max} 
    \approx \bar R_{*} \left( \frac{\lambda_{F \to T}}{3}  \right)^{-\frac{1}{3}}.
\ee

Finally, we consider that expanding TVBs may nucleate inside the FVBs also after inflation. The corresponding nucleation rate can be estimated as~\cite{Linde:1981zj}
\be
    \Gamma  \propto  \left(\frac{S}{2\pi}\right)^2 e^{-S}, \qquad
    S \approx \frac{27 \pi^2 \sigma^4}{2 \Delta V^3} \approx \frac{\pi^2}{24} \frac{\kappa^2 v^4}{\Delta V} \,,
\ee
where we used the dimensional prefactor $(3\sigma/\Delta V)^{-4}$ to describe the initial radius of a critical size TVB (see Eq.~\eqref{eq:TVB_crit}). We have checked that our results are not very sensitive to this choice. The probability that TVBs have not nucleated in an expanding FVB with comoving volume $V(R_0)$ evolves as $\dot P_{\rm nucl,\omega} = -a^3 V(R_0) \Gamma P_{\rm nucl,\omega}$. Assuming that, after inflation, the Universe instantaneously transitions to an era dominated by a perfect fluid with equation of state parameter $\omega$, so that $H \propto a^{-3(1+\omega)/2}$, we estimate that each FVB re-enters the horizon when the latter reaches the size
\be
    H_{0} = \frac{1}{a_{0}R_0} = \Hinf \left(\frac{R_{\rm inf}}{R_0}\right)^{\frac{3 (1+\omega)}{1+3 \omega}}
\ee
and the probability that the entire FVB remains in the false vacuum is
\be
    P_{\rm nucl,\omega} = \exp\left( -\frac{8\pi \Gamma}{9(1+\omega) H_{0}^4} \right) \,.
\ee
As before, in order to account for the destruction of FVBs due to tunneling before the collapse, we make the replacement $\td n_B(R_0) \to P_{\rm nucl}(R_0) \td n_B(R_0)$.

In the following, we consider only the cases of RD ($\omega = 1/3$) and MD ($\omega = 0$) backgrounds, obtained respectively with a prompt or delayed decay of a sufficiently massive inflaton field. Deviations from these regimes may occur depending on the inflaton potential, as in the case of a coherently oscillating inflaton~\cite{Turner:1983he}. However, such configurations are generally short lived due to the rapid fragmentation of the inflaton,  which yields a RD or MD epoch at most within $\mathcal{O}(1)$ $e$-folds~\cite{Lozanov:2016hid,Lozanov:2017hjm,Tomberg:2021bll}. For simplicity, we also assume that $\omega$ is constant in the simulations of collapsing FVB presented below.

\section{Collapse of false vacuum bubbles}
\label{sec:FVBs}

To study the collapse of the false vacuum domains produced by the spectator field, we solve the Klein-Gordon equation numerically on a lattice. For simplicity, we simulate only $O(3)$ symmetric FVBs using spherical comoving coordinates $r$, $\theta$, $\varphi$ and conformal time $\eta$, in an expanding background $\td s^2 = a^2 [-\td \eta^2 + \td r^2 + r^2 (\td\theta^2 + \sin^2\theta \,\td \varphi^2)]$ where $a$ is the scale factor. We neglect the effect of the scalar field density on the metric, so the dynamics can be fully captured by the Klein-Gordon equation
\be
    \partial_\eta^2 \phi + \frac{2\partial_\eta a}{a}\partial_\eta \phi - \partial_r^2 \phi - \frac{2}{r} \partial_r \phi = -a^2\frac{\td V}{\td \phi} \,.
\ee
In the RD case the scale factor is given by $a = a_0^2 H_0 \eta$, where $H_0$ is the Hubble rate at $a=a_0$, whereas in the MD case $a = a_0^3 H_0^2 \eta^2/4$. For the numerical simulations we define dimensionless variables $\chi \equiv \phi/v$, $x'^{\mu} \equiv a_0 H_0 x^{\mu}$, so that
\be
    \partial_{\eta'}^2 \chi + \frac{2\beta}{\eta'} \partial_{\eta'} \chi - \partial_{r'}^2 \chi - \frac{2}{r'} \partial_{r'} \chi = - \frac{\eta'^{2\beta}}{\beta^4} \frac{\Delta V}{v^2H_0^2} \frac{\td \tilde V}{\td \chi} \,,
\ee
where $\beta=1$ for RD and $\beta=2$ for MD, and $\tilde V(\chi) = V(\chi v)/\Delta V$. We parametrize the initial field profile as
\be \label{eq:phiin}
    \frac{\phi_0}{v} =  \frac{1}{2} \left[1 - \tanh \left(2\frac{r-R_0}{wR_0}\right) \right] \, ,
\ee
with $w$ being the relative width of the FVB wall, $R_0$ the initial radius of the FVB and $a_0$, chosen such that $a_0 H_0 R_0=1$, that is, at the time the FVB re-enters the horizon the Hubble rate is $H_0$. In this parametrization the numerical problem involves three free parameters: $\kappa$, that determines the height of the potential energy barrier, $w$, that determines the initial width of the bubble wall, and $\Delta V/v^2H_0^2$, determined by the potential energy difference and affecting the separation of the minima and the initial bubble size.

\begin{figure}
\centering
\includegraphics[width=0.9\textwidth]{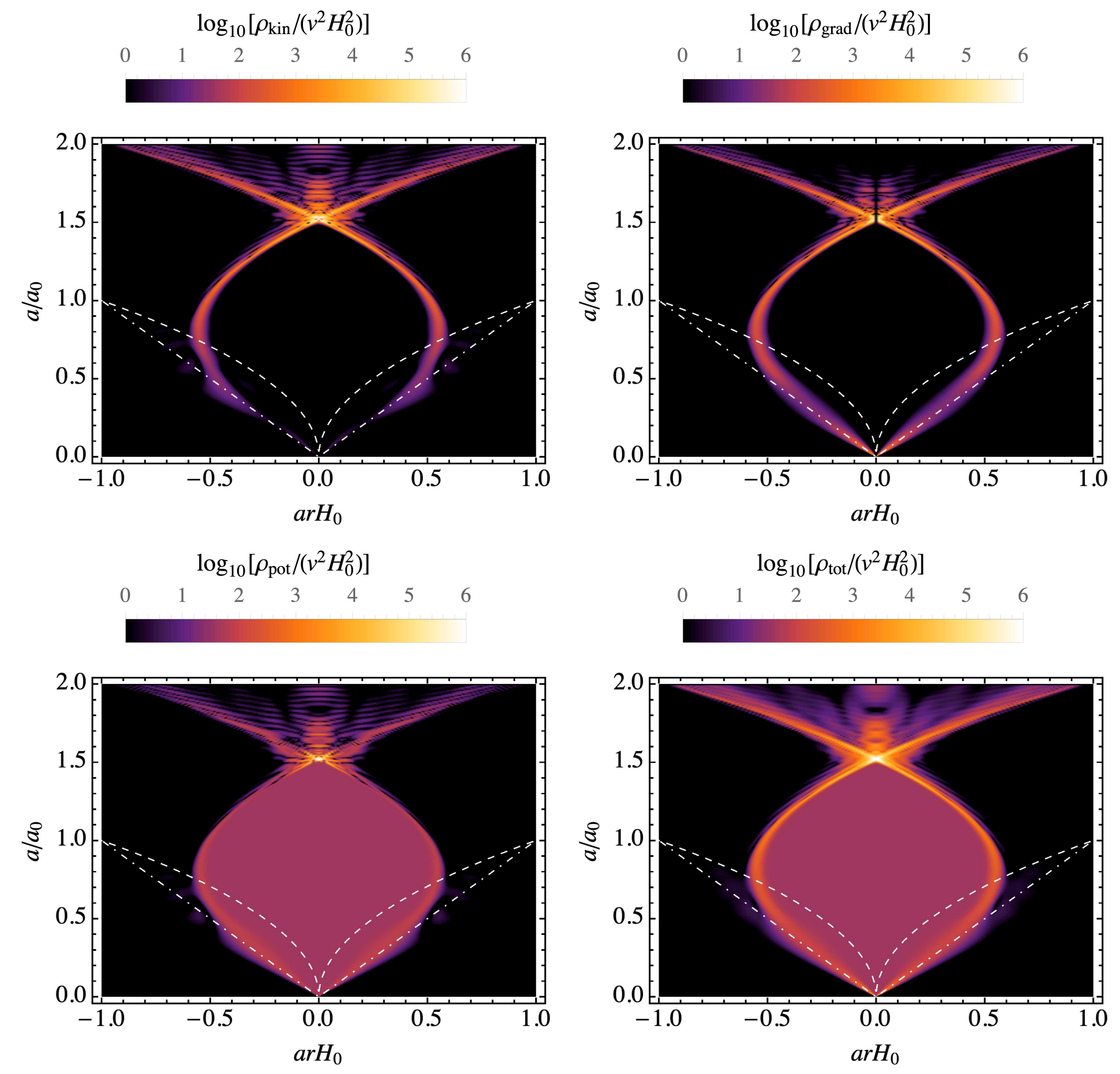}
\caption{Time evolution of the energy density components of a collapsing FVB as a function of the scale factor $a$. The dashed curves indicate the horizon radius, $H_0/H$, and the dot-dashed lines correspond to $|r| = 1/(a_0H_0)$. Here $\kappa=20.0$, $\sqrt{\Delta V}/(v H_0)=40.0$ and $w=0.2$.}
\label{fig:rho}
\end{figure}

The relevant quantities for BH formation in the FVB collapse are the mass contained in a spherical region of comoving radius $r$ around the center of the bubble,
\be
    m(r) = 4\pi \,a^3 \int_0^r \!\td r' \,r'^2 \rho_{\rm tot}(r') \,,
\ee 
and its compactness,
\be
    C(r) = \frac{m(r)}{8\pi \mpl^2 a r} \,.
\ee
The total energy density of the bubble is a sum of kinetic, gradient and potential energies, $\rho_{\rm tot} = \rho_{\rm kin} + \rho_{\rm grad} + \rho_{\rm pot}$, given by
\be
    \rho_{\rm kin} = \frac{(\partial_\eta \phi)^2}{2a^2}\,, \quad 
    \rho_{\rm grad} =  \frac{(\partial_r \phi)^2}{2a^2}\,, \quad
    \rho_{\rm pot} = V(\phi) \,.
\ee
In order to determine whether the FVBs collapse into BHs we employ the hoop conjecture~\cite{Thorne1995}, that is, we assume that a BH is formed whenever the compactness of the FVB exceeds the compactness of a BH, $C>1/2$. A complementary approach was adapted in Refs.~\cite{Deng:2016vzb,Deng:2017uwc}, where the Einstein field equations were solved considering the scalar field bubbles only in the thin-wall limit.

\begin{figure}
\centering
\includegraphics[width=\textwidth]{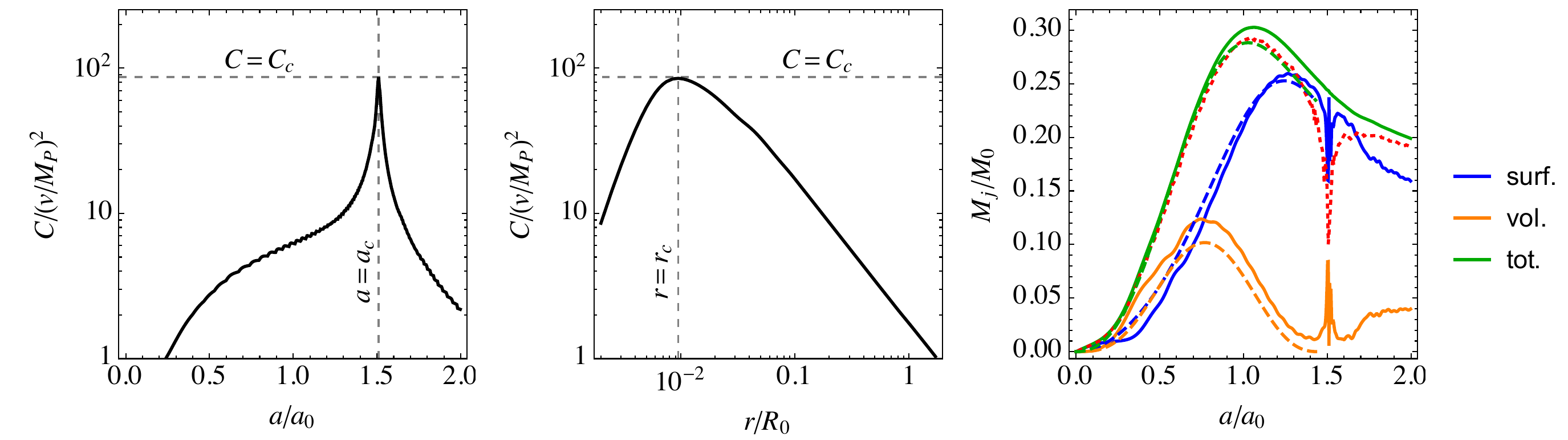}
\caption{\emph{Left panel:} The evolution of the maximal compactness of the FVB. \emph{Middle panel:} The compactness of mass contained in a comoving radius $r$ at $a=a_c$, maximized for $r = r_c$. \emph{Right panel:} The solid curves show the evolution of the total mass (green), the surface mass (blue) and the volume mass (orange) of the FVB in the lattice simulation. For comparison, the corresponding dashed curves show the evolution obtained in the thin-wall approximation. The red dotted curve shows the mass of the most compact region obtained from the lattice simulation.}
\label{fig:comp}
\end{figure}

In Fig.~\ref{fig:rho} we show the evolution of the energy density components for a benchmark case with $\kappa=20.0$, $\sqrt{\Delta V}/(v H_0)=40.0$ and $w=0.2$. We see that when the FVB radius is larger than the Hubble horizon, indicated by the white dashed curves, the physical FVB radius increases. At $a\ll a_0$ the comoving radius of the FVB is constant. As the Hubble radius reaches the FVB size, the bubble starts to shrink and the potential energy is transferred to the kinetic and gradient energies of the wall. Eventually, as the FVB radius approaches zero, very high energy density can be reached at the collision point. The maximal compactness of the FVB increases during the collapse as shown in the left panel of Fig.~\ref{fig:comp}. The middle panel shows instead the compactness at the end of the collapse as a function of the radius $r$, indicating that the maximal compactness is reached at a finite radius.

In the right panel of Fig.~\ref{fig:comp} we present the evolution of the surface and volume masses of the FVB,
\be
     M_{\rm surf} = 4\pi \,a^3 \int \!\td r \,r^2 \left[\rho_{\rm kin}(r) + \rho_{\rm grad}(r)\right] \,, \quad 
    M_{\rm vol} = 4\pi \,a^3 \int \!\td r \,r^2 \rho_{\rm pot}(r) \,,
\ee 
as well as of their sum $M_{\rm tot} = M_{\rm surf} + M_{\rm vol}$. We also show the surface and volume masses computed in the thin-wall limit for a FVB obtained by evolving the comoving FVB radius\footnote{The gravitational self-interaction of the FVB can be included via the shift $\Delta V \to \Delta V + (3/4) \sigma^2/\mpl^2$~\cite{Deng:2017uwc,Deng:2020mds}. The contribution can be neglected when $\sigma^2 \ll \Delta V \mpl^2$. Eq.~\eqref{eq:sigma} then implies that $v \ll \sqrt{24/\kappa}\mpl$, thus the condition is expected to hold for sub-Planckian field values.}
\be\label{eq:TW_eom}
    \partial_\eta^2 R + \left( \frac{2}{R} +3\frac{\partial_\eta a}{a} \partial_\eta R \right) (1 - \partial_\eta R^2) 
    = - a\frac{\Delta V}{\sigma} (1-\partial_\eta R^2)^{3/2} \,,
\ee
where
\be\label{eq:sigma}
    \sigma 
    \equiv \int_{\phi_c}^v \td \phi \,\sqrt{2(V(\phi)-\Delta V)}
    \approx  \frac{v \sqrt{2\Delta V \kappa}}{6(1+\sqrt{10}/\kappa)^{2}}
\ee
denotes the surface tension of the bubble and $\phi_c = v (1+\sqrt{4+\kappa})/(3+\kappa)$ is the field value between the minima satisfying $V(\phi_c) = V(v) \equiv \Delta V$ (see Fig.~\ref{fig:pot}). The approximation for $\sigma$ was obtained by interpolating between the $\kappa \to 0$ and $\kappa \to \infty$ asymptotics, which read $\sigma \sim v \sqrt{2\kappa \Delta V}/6$ and $\sigma \sim v \sqrt{2\Delta V} \kappa^{5/2}/60$, respectively. In this approximation, the surface, volume and total masses are\footnote{We can neglect the mass of the radiation enclosed within the FVB volume and self-gravity, which would both contribute to the Misner-Sharp mass of the bubble~\cite{Deng:2020mds}.}
\be
    M_{\rm surf} = 4\pi\sigma \frac{a^2 R^2}{\sqrt{1-\partial_\eta R^2}} \,, \quad 
    M_{\rm vol} = \frac{4\pi}{3} a^3 R^3 \Delta V\,, \quad
    M_{\rm tot} = M_{\rm surf} + M_{\rm vol} \,.
\ee
Although the thin-wall approximation estimates the evolution of the FVB well, it does not allow to resolve the maximal compactness reached in the collapse. For the latter we use the lattice simulation.

\begin{figure}
\centering
\includegraphics[width=0.9\textwidth]{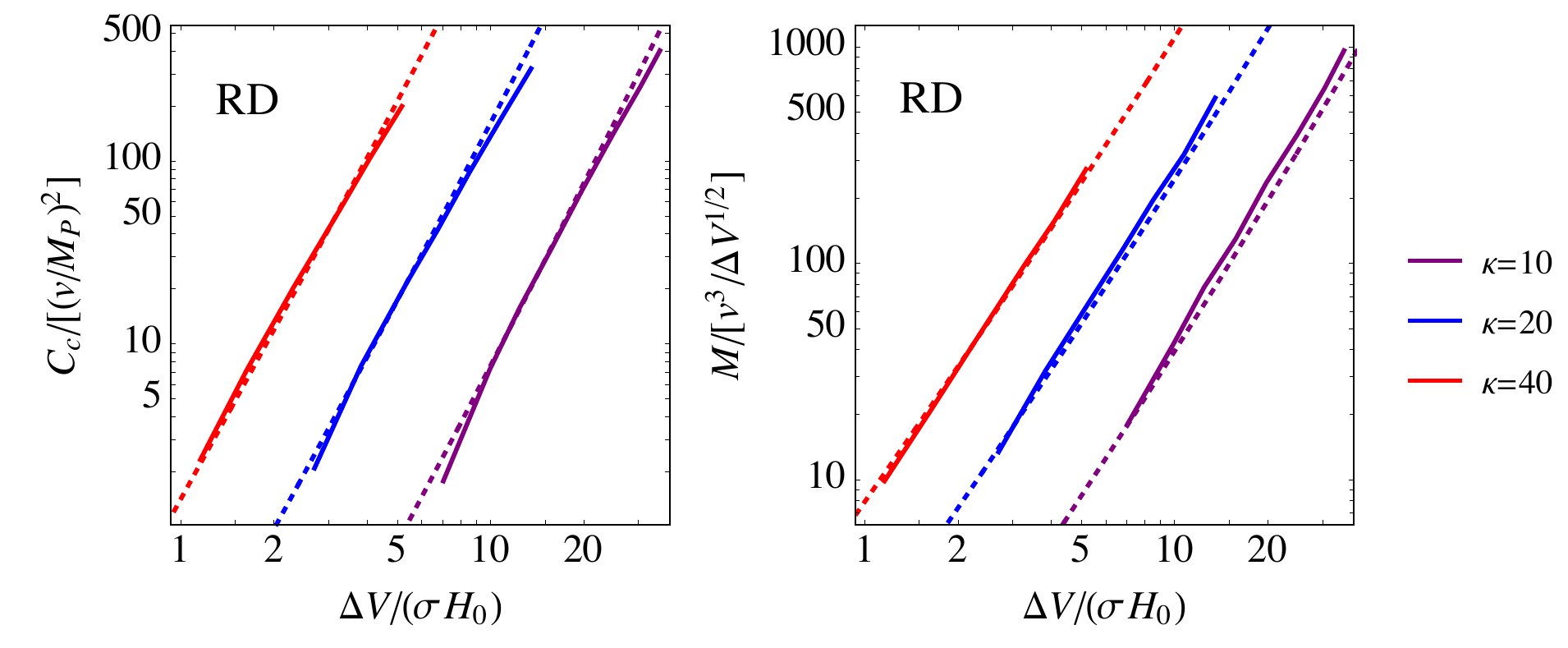}
\includegraphics[width=0.9\textwidth]{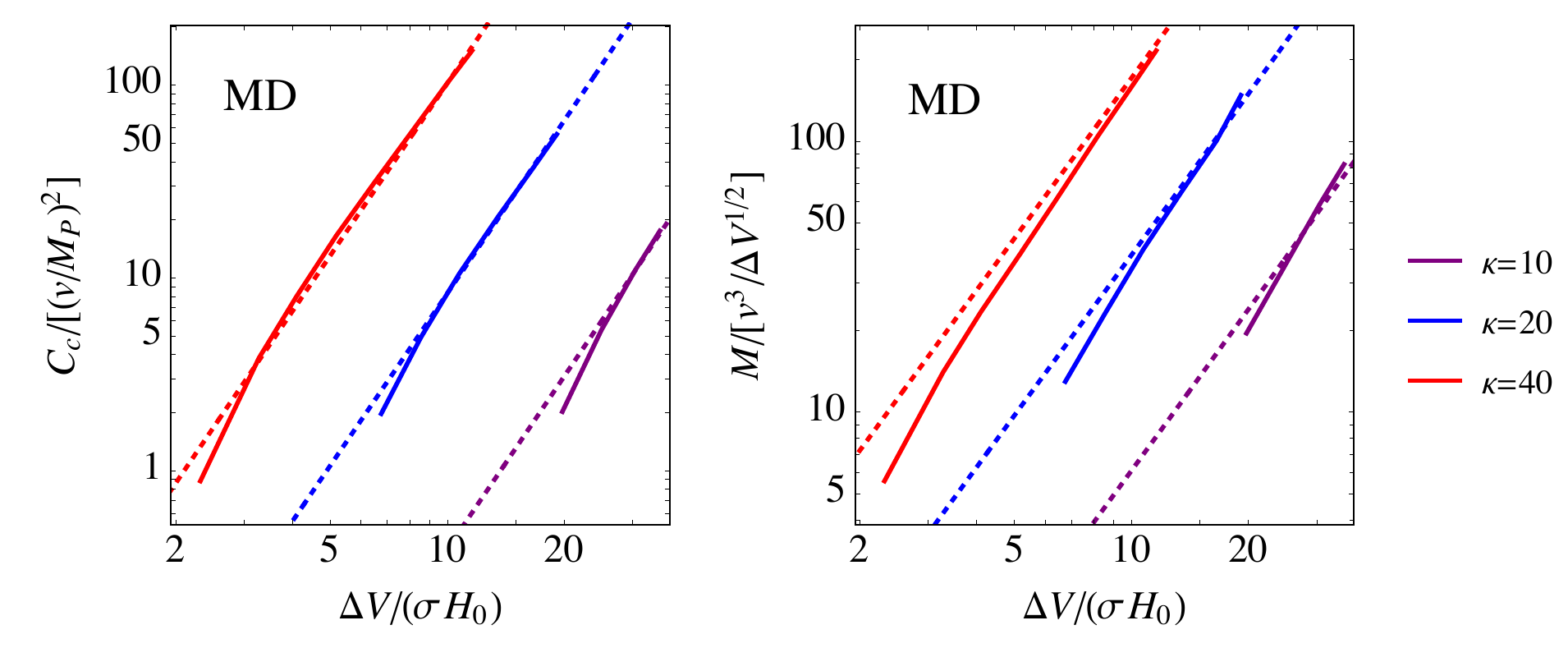}
\caption{Maximal compactness \emph{(left panels)} and  mass \emph{(right panels)} of the maximally compact region as a function of $\Delta V/(\sigma H_0)$ for different values of $\kappa$. The upper and lower panel correspond to a RD or MD background, respectively. The dotted lines show the fits~\eqref{eq:CMfitRD}~\eqref{eq:CMfitMD} obtained for $w=0.2$, but the results remain substantially the same for other $\mathcal{O}(0.1)$ values of $w$.}
\label{fig:R0dep}
\end{figure}

\begin{figure}
\centering
\includegraphics[width=0.6\textwidth]{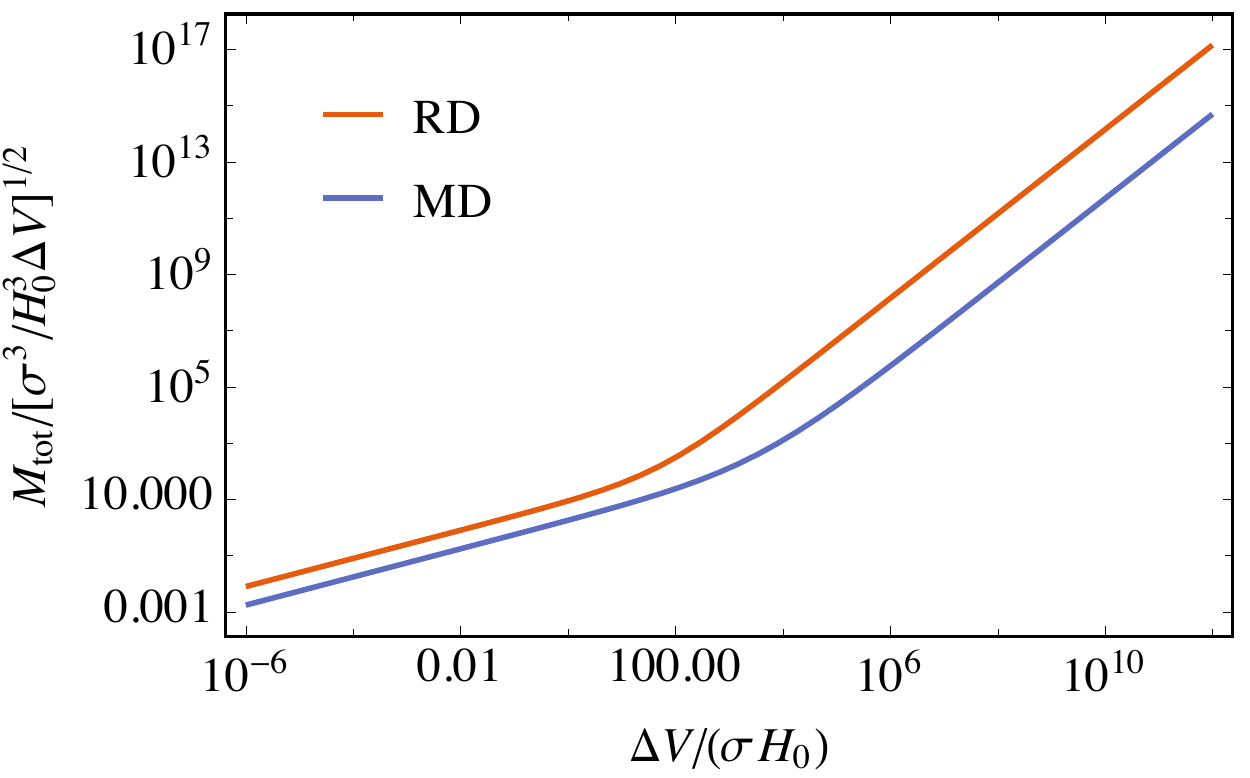}
\caption{The total mass of the FVB at the end of the collapse obtained from the thin-wall approximation in RD and MD backgrounds.}
\label{fig:Mtot}
\end{figure}

In Fig.~\ref{fig:R0dep} we show by the solid curves the maximal compactness $C_c$ and the mass $M_c$ reached in the most compact region obtained from the lattice simulations using a fixed initial bubble wall width $w=0.2$. We have checked that the result does not significantly change for different $\mathcal{O}(0.1)$ values of $w$. We find that the maximal compactness $C_c$ and the corresponding mass $M_c$ can be approximated by
\be \label{eq:CMfitRD}
    C_c \approx 0.06\, \frac{\sigma M_{\rm tot}(a_c)}{v^2 \mpl^2} \,, \qquad   
    M_c \approx 0.5 M_{\rm tot}(a_c) \,.
\ee
in RD and by
\be\label{eq:CMfitMD}
    C_c \approx 0.02\, \frac{\sigma M_{\rm tot}(a_c)}{v^2 \mpl^2} \,, \qquad   
    M_c \approx 0.5 M_{\rm tot}(a_c) \,.
\ee
in MD. These relations are plotted in Fig.~\ref{fig:R0dep} with dashed curves. In order to extend these results to regions where the lattice simulation cannot be used due to numerical limitations, we estimate the total mass at the collapse moment, $M_{\rm tot}(a_c)$, using the thin-wall results at the collapse moment, $a=a_c$. We find that it is well approximated by a broken power-law
\be\label{eq:MtotfitRD}
    M_{\rm tot}(a_c) \approx 0.14\, \frac{\Delta V}{H_0^3} \left[1 + 14.8 \left(\frac{\sigma H_0}{\Delta V}\right)^{\!2/3} \right]^{3/2} 
\ee
in RD and 
\be\label{eq:MtotfitMD}
    M_{\rm tot}(a_c) \approx 4.9\times 10^{-4}\, \frac{\Delta V}{H_0^3} \left[1 + 59.7 \left(\frac{\sigma H_0}{\Delta V}\right)^{\!1/2} \right]^{2}
\ee
in MD, as presented in Fig.~\ref{fig:Mtot}. For small values of $H_0$ the volume mass comprises most of the total mass of the FVB and the total mass scales as $M_{\rm tot} \propto H_0^{-3}$, while for large values the surface mass dominates and the total mass scales as $M_{\rm tot} \propto H_0^{-2}$. The behavior changes at $M_{\rm tot} \simeq 7.3\times 10^4 \sigma^3/\Delta V^2$ in RD and at $M \simeq 8.9\times 10^7 \sigma^3/\Delta V^2$ in MD. 

Finally, the critical size below which the TVBs will collapse can be estimated by using the thin-wall approximation~\eqref{eq:TW_eom} with a positive sign on the right hand side of the equality. The bubble contracts when $\partial_\eta R < 0$ and is guaranteed to continue contracting as long as $\partial_\eta^2 R \leq 0$ in Eq.~\eqref{eq:TW_eom}. By neglecting the expansion after the TVB has entered horizon, we see from Eq.~\eqref{eq:TW_eom} that it continues contracting when $\Delta V/\sigma \lesssim  2/(a_0 R_0) = 2H_0$. Consequently, TVBs indeed collapse as long as they are sufficiently small. This rough analytic estimate was confirmed numerically by solving Eq.~\eqref{eq:TW_eom} in an expanding background. We find that if
\be\label{eq:TVB_crit}
    \frac{\Delta V}{\sigma H_0} < 1.6 \quad \mbox{(RD)}, 
    \qquad \qquad
    \frac{\Delta V}{\sigma H_0} < 2.7 \quad \mbox{(MD)},
\ee
then all TVBs, possibly nucleated inside the FVB during inflation, collapse after horizon re-entry and do not destroy the enclosing FVB. From the above relations we can consequently estimate the maximal initial comoving radius~\eqref{eq:R_max} of the FVBs that may form PBHs.

\section{Primordial black holes}
\label{sec:PBHs}

Knowing the comoving number density of the FVBs, $\td n_{B}/\td R_0$, we  estimate the resulting comoving number density of PBHs with a given mass $M$ as
\bea
    \frac{\td n_{\rm PBH}}{\td M} 
&    = \int \td R_0 \, \theta(C(R_0)-1/2) \delta(M-M_{\rm PBH}(R_0)) \frac{\td n_B}{\td R_0} \,
    \\
&    = \theta(M-M_{\rm min}) \left(\frac{\td M}{\td R_0}\right)^{-1} \!\left. \frac{\td n_B}{\td R_0} \, \right|_{M = M_{\rm PBH}(R_0)} \,,
\eea
where $M_{\rm PBH}(R_0)$ is the mass of the PBH formed from a FVB of initial comoving radius $R_0=1/(a_0 H_0)$, as given in Eq.~\eqref{eq:CMfitRD} or~\eqref{eq:CMfitMD} depending on whether the collapse takes place in RD or MD. The condition $C > 1/2$ for PBH formation implies a minimal initial comoving FVB radius $R_{\rm min}$ that corresponds to the minimal PBH mass, $M_{\rm min} = M_{\rm PBH}(R_{\rm min})$. For $\kappa\gg 1$ we have
\be\label{eq:Mmin}
    M_{\rm min}
    \approx 17.8 M_{\rm P}^2 v/\sqrt{\kappa \Delta V} \quad \mbox{(RD)},  
    \qquad
    M_{\rm min}
    \approx 53.0 M_{\rm P}^2 v/\sqrt{\kappa \Delta V} \quad \mbox{(MD)}.
\ee 
The condition~\eqref{eq:cond_V2} guarantees that the PBHs have super-planckian masses. In fact, with the latest Planck constraint $\Hinf < 2.5\times 10^{-5}\mpl$~\cite{Planck:2018jri} we obtain $M_{\rm min} \gtrsim 25 \mpl^2/\Hinf > 4 \, \rm g$, in RD, and $M_{\rm min} \gtrsim 76 \mpl^2/\Hinf > 13 \, \rm g$ in MD. This further implies, that FVBs cannot collapse into BHs immediately after inflation as $M_{\rm tot}(H_0 = \Hinf) \lesssim \Hinf \ll \mpl$.

The total PBH energy density is
\be\label{eq:rho_PBH}
    \rho_{\rm PBH} 
    = \int \td M \, M \frac{\td n_{\rm PBH}}{\td M} 
    = \int_{R_{\rm min}}^\infty \td R_0 \,M \left. \frac{\td n_B}{\td R_0} \right|_{M = M_{\rm PBH}(R_0)} \,,
\ee
whereas the PBH mass function reads
\be
    \psi(M) 
    \equiv \frac{M}{\rho_{\rm DM}}\frac{\td n_{\rm PBH} }{\td M} \,,
\ee
and is normalized to the fraction of DM in PBH: $\int \td M \, \psi(M) = f_{\rm PBH} \equiv \rho_{\rm PBH}/\rho_{\rm DM}$. Through Eq.~\eqref{eq:nBnoneq} we find that $\psi(M)$ is a broken power-law that interpolates between $\psi\propto M^{-3/4}$, at low masses, and $\psi\propto M^{-1/2}$ at high masses. Unless $\kappa v \gtrsim 0.14 M_{\rm P}$ in RD or $\kappa v \gtrsim 0.007 M_{\rm P}$ in MD, the minimal PBH mass is so large that $\psi\propto M^{-1/2}$.

Assuming $\kappa \gg 1$, if $\kappa v \ll 0.14 M_{\rm P}$ in RD or $\kappa v \ll 0.007 M_{\rm P}$ in MD, the minimal PBH mass is so large that the surface mass contribution to the total FVB mass can be neglected in Eqs.~\eqref{eq:MtotfitMD} and \eqref{eq:MtotfitRD}. Then, $M_{\rm PBH} = 0.070\Delta V/H_0^3$ in RD and $M_{\rm PBH} = 2.5\times 10^{-4} \Delta V/H_0^3$ in MD. In this case, the present PBH abundance can be approximated by
\bea\label{eq:f_RD+MD}
    f_{\rm PBH} 
&    \approx 14 
    \sqrt{\frac{\Delta V}{\Hinf^4}} e^{-\frac{\pi^2}{6} \kappa \frac{\Delta V}{\Hinf^4}}
    \left(\frac{M_{\rm min}}{10^{12}\,{\rm g}} \right)^{\!-\frac32} \left( \sqrt{\frac{M_{\rm max}}{M_{\rm min}}} -1 \right)
    \qquad \qquad \qquad \, \ \mbox{(RD)}, 
\\
    f_{\rm PBH} 
&   \approx 3.6 
    \sqrt{\frac{\Delta V}{\Hinf^4}} e^{-\frac{\pi^2}{6} \kappa \frac{\Delta V}{\Hinf^4}}
    \left(\frac{M_{\rm min}}{10^{12}\,{\rm g}} \right)^{\!-\frac32} \left[ \left(\frac{M_{\rm max}}{M_{\rm min}}\right)^{\!\frac23} - 1 \right] 
    \left(\frac{M_{\rm min}}{M_{\rm MD}}\right)^{\!\frac16} 
    \quad \ \mbox{(MD)}, 
\eea
where $M_{\rm MD}$ denotes the mass of the BH formed by a FVB that would re-enter the horizon at the end of MD. We stress that these expressions ignore evaporation through Hawking radiation~\cite{Hawking:1974rv,Hawking:1974sw}, which would prevent populations characterized by a large number of light PBHs from resulting in significant values of $f_{\rm PBH}$ at present time.

\begin{figure}
\centering
\includegraphics[height=0.3\textwidth]{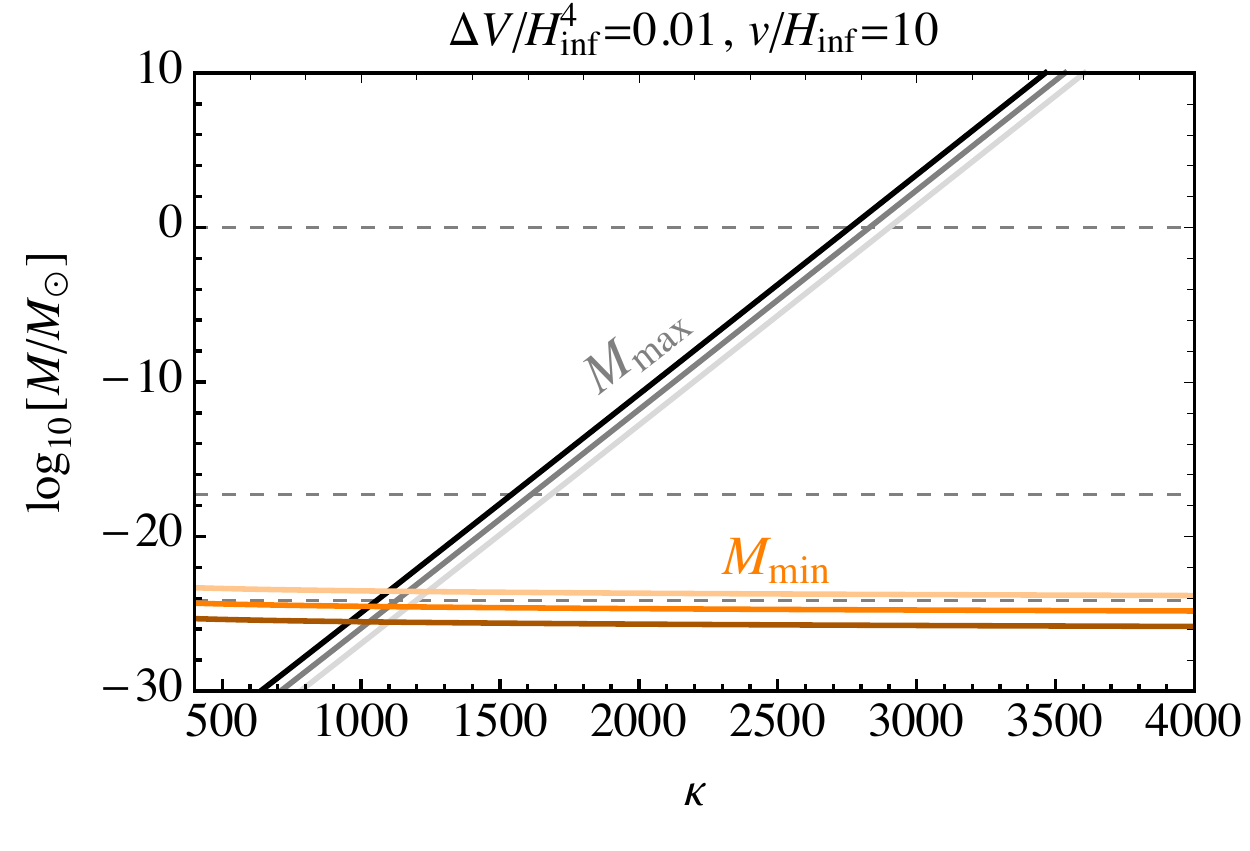}
\includegraphics[height=0.3\textwidth]{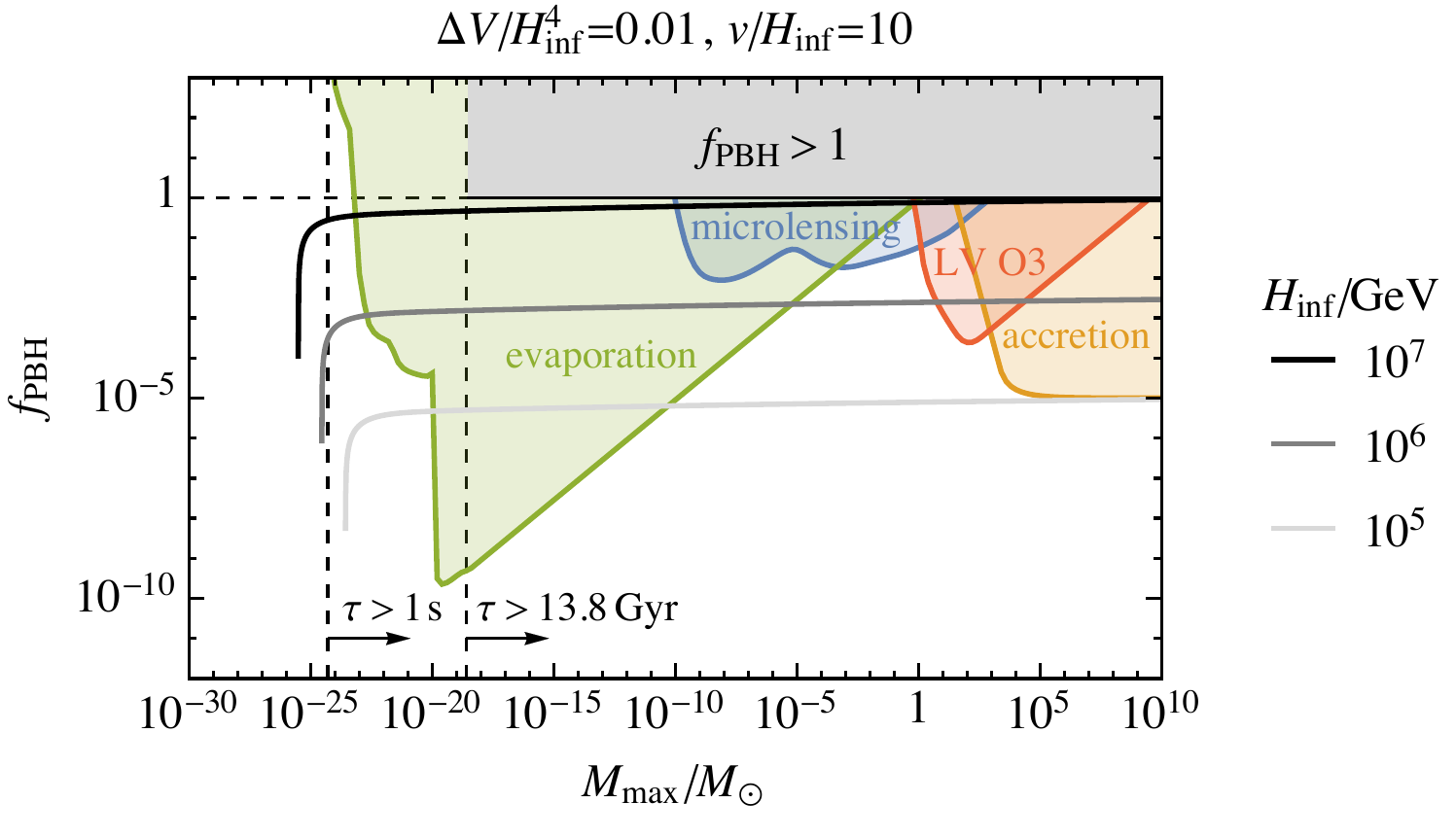}
\caption{\emph{Left panel:} $M_{\rm max}$ and $M_{\rm min}$ as a function of $\kappa$ for fixed $\Delta V/\Hinf^4$, $v$ and $v/\Hinf$ and different $\Hinf$. \emph{Right panel:} Constraints on the PBH abundance for the approximate mass function~\eqref{eq:psi_simp} as a function of $M_{\rm max}$. The gray lines show the predicted abundance for the same parameters as in the left panel. The parameter $\kappa$ varies along these lines. The vertical dashed lines indicate the PBH masses corresponding to evaporation after BBN and the present.}
\label{fig:PBH}
\end{figure}

Eqs.~\eqref{eq:f_RD+MD} imply that the mechanism results only in a small abundance of PBHs when $M_{\rm min}$ exceeds the evaporation limit of $10^{17}$\,g. In fact, suppose that these PBHs occupy the currently unconstrained mass window, which sets $M_{\rm min} = 10^{17} \rm g$ and $M_{\rm max} = 10^{23} \rm g$, and momentarily regard these parameters as independent. The condition~\eqref{eq:cond_V2} implies that $\Delta V/\Hinf^4 \gsim 12/(\pi^2 \kappa)$, so the abundance is suppressed by the exponential prefactor and results in $f_{\rm PBH} \lesssim 10^{-4}/\sqrt{\kappa}$ on the considered mass window. Larger values $f_{\rm PBH} = \mathcal{O}(1)$ can be reached only with a $M_{\rm max}$ close to, or higher, than the solar mass range. However, in this case $f_{\rm PBH} = 1$ is ruled out by microlensing observations~\cite{Niikura:2017zjd,Smyth:2019whb,Tisserand:2006zx,Niikura:2019kqi,Griest:2013aaa,Macho:2000nvd,Allsman:2000kg}, the LIGO-Virgo observations~\cite{Raidal:2017mfl,Ali-Haimoud:2017rtz,Raidal:2018bbj,Vaskonen:2019jpv,DeLuca:2020qqa,Hutsi:2020sol} or by CMB limits on accretion~\cite{Ricotti:2007au,Horowitz:2016lib,Ali-Haimoud:2016mbv,Poulin:2017bwe,Hektor:2018qqw,Serpico:2020ehh}. Finally, as the mass function is a relatively flat power law, it is not possible to produce a low abundance of $M \ll 10^{17} \rm g$ PBHs, avoiding the evaporation constraints, and have at the same time a non-evaporating PBH population that constitutes all of DM. In particular, as can be seen from Fig.~\eqref{fig:PBH}, the asteroid mass window for PBH DM is closed by the evaporation constraints~\cite{Carr:2009jm,Acharya:2020jbv}. On the basis of the above argument we therefore conclude that PBHs formed by collapsing spectator field FVBs can constitute only a subdominant DM component.

The maximal mass is determined mainly by formation of critical size TVBs during inflation, as the condition~\eqref{eq:cond_V2} implies that nucleation of TVBs during RD is subdominant. Eq.~\eqref{eq:R_max} then gives
\bea\label{eq:Mmax}
    M_{\rm max} 
    &\approx 
    1.5 \frac{\kappa^2 v^7}{\Hinf^6} \left(\frac{\kappa\Delta V}{\Hinf^4}\right)^{-\frac{5}{2}} e^{\frac{\pi^2}{3} \kappa \frac{\Delta V}{\Hinf^4}} \\
    &=
    2.7 \times 10^{-11} {\rm g} \, \kappa^{-\frac{1}{2}}  
    \left[\frac{\Hinf}{10^{13}\GeV} \right] 
    \left[\frac{v}{\Hinf} \right]^{7}  
    \left[\frac{\Delta V}{\Hinf^4}\right]^{-\frac{5}{2}} e^{\frac{\pi^2}{3} \kappa \frac{\Delta V}{\Hinf^4}} \,.
\eea
Heavy PBHs can thus be produced if the exponential factor is large, implying that the nucleation of expanding TVBs inside the FVBs is correspondingly strongly suppressed. In the presence of a post-inflationary MD epoch, however, the maximal size of the FVB can be affected by the MD to RD transition if the radiation bath couples to the spectator field. Thermal corrections to the spectators potential can lift the second minimum and allow the field to roll to the true vacuum, thus destroying the FVBs that would otherwise re-enter after the transition. In this case, $M_{\rm MD}$ constitutes the maximal mass of the BHs.  

Plugging the mass limits \eqref{eq:Mmin} and \eqref{eq:Mmax} into Eq.~\eqref{eq:f_RD+MD} gives that, in RD,
\be\label{eq:f2_RD}
    f_{\rm PBH} \approx 
    2.0 \times 10^{11} \, \kappa^{\frac34}
    \left[\frac{\Hinf}{10^{13}\GeV} \right]^{\frac{5}{2}} 
    \left[\frac{v}{\Hinf} \right]^{\frac{3}{2}} 
    \left[\frac{\Delta V}{\Hinf^4} \right]^{\frac{1}{4}} \,,
\ee
if $M_{\rm max} \gg M_{\rm min}$. The exponential factors in \eqref{eq:f_RD+MD} and \eqref{eq:Mmax} cancel out at leading order in $\kappa$. Notice that the condition~\eqref{eq:cond_V2} implies a lower bound for the last two factors in \eqref{eq:f2_RD}, therefore $f_{\rm PBH}$ can be reduced only by decreasing $\Hinf$. 

The mass function for $M_{\rm max} \gg M_{\rm min}$ is approximately of the form
\be\label{eq:psi_simp}
    \psi(M) \approx  \frac{f_{\rm PBH}}{\sqrt{2 M M_{\rm max}}} \,\theta(M_{\rm max}-M) \,.
\ee
Fig.~\ref{fig:PBH} shows the strongest bounds it must satisfy, arising from PBH evaporation during BBN and non-observation of extragalactic gamma-ray background from Hawking radiation~\cite{Carr:2009jm,Acharya:2020jbv}, microlensing from Subaru/HSC~\cite{Niikura:2017zjd,Smyth:2019whb}, EROS~\cite{Tisserand:2006zx}, OGLE~\cite{Niikura:2019kqi}, Kepler~\cite{Griest:2013aaa} and MACHO~\cite{Macho:2000nvd,Allsman:2000kg}, LIGO-Virgo observations~\cite{Raidal:2017mfl,Ali-Haimoud:2017rtz,Raidal:2018bbj,Vaskonen:2019jpv,DeLuca:2020qqa,Hutsi:2020sol}, and limits on accretion~\cite{Ricotti:2007au,Horowitz:2016lib,Ali-Haimoud:2016mbv,Poulin:2017bwe,Hektor:2018qqw,Serpico:2020ehh}. These constrains were adapted to the extended mass function~\eqref{eq:psi_simp} by using the method derived in Ref.~\cite{Carr:2017jsz}. 

As discussed above, the PBH population produced by our mechanism is generally unsuitable to explain the observed DM abundance in the allowed asteroid mass window. Still, we find that a significant PBH abundance can be generated in the mass range relevant for LIGO-Virgo observations or for seeding supermassive BHs. As seen from Fig.~\ref{fig:PBH}, this can be achieved for example with $\Delta V/\Hinf^4 = 0.01$, $v/\Hinf = 10$, $\kappa \simeq 3000$ and $\Hinf/{\rm GeV} \simeq 10^5 - 10^6$.

Another interesting phenomenological consequence is a possible transient MD epoch in the early Universe, where the energy budget of the Universe is briefly taken over by evaporating PBHs. In fact, from Eq.~\eqref{eq:f_RD+MD} we see that the produced abundance may be much larger than unity, $f_{\rm PBH} \gg 1$. Therefore, if initially in a RD regime, the corresponding PBH population would come to dominate the energy budget at about $a \approx a_{\rm eq}/f_{\rm PBH}$, where $a_{\rm eq}$ denotes the the matter-radiation equality. In terms of temperature, the onset of this transient MD epoch is at $T \gtrsim 1\,\eV \, f_{\rm PBH}$, corresponding to an age of the Universe of $t_{\rm dom} = 10^{12}\, {\rm s} / f_{\rm PBH}^2$. For the parameters used in Eq.~\eqref{eq:f2_RD}, for instance, PBHs start to dominate the energy density close to the electroweak phase transition epoch. 

The transient MD regime can occur if the PBHs evaporate on timescales larger than $t_{\rm dom}$. In regard of this, because the evaporation time of a of a non-spinning BH with initial mass $M$~\cite{Page:1976df}
\be
    \tau(M) = 0.4\,{\rm s} \left(M/10^9\,{\rm g}\right)^3,
\ee
is generally much larger than the age of the Universe at the time of PBH formation, we can neglect the time of formation. Moreover, since the obtained PBH mass function \eqref{eq:psi_simp} is top-heavy, we can estimate the evaporation time of the PBH population as $\tau(M_{\rm max})$, obtaining that the Universe undergoes a transient MD period if
\be
    \tau(M_{\rm max}) \gg t_{\rm dom} \,.
\ee
Furthermore, a consistent cosmology requires that any possible transient MD period completes well before the onset of BBN~\cite{Allahverdi:2020bys}, implying that $\tau(M_{\rm max}) \lesssim 0.1 {\rm s}$ or $M_{\rm max} \lesssim 10^9\,{\rm g}$. This bound neglects the gravitational wave production by the evaporating PBHs, which will contribute to the energy density of radiation and thus further strengthen the BBN constraints~\cite{Dong:2015yjs,Domenech:2020ssp}.

\section{Conclusions}
\label{sec:concl}

We studied the production of false vacuum bubbles and primordial black holes due to fluctuations of a light spectator field in an asymmetric polynomial potential with two non-degenerate minima. Using the stochastic formalism, we showed that infrequent fluctuations can populate the minimum corresponding to a false vacuum state, thereby resulting in the formation of false vacuum bubbles. These domains expand during inflation and collapse after re-entering the horizon in the subsequent evolution of the Universe.

We analyzed the collapse of the false vacuum bubbles by means of a dedicated lattice simulation and compared our findings with analytical estimates that use the thin-wall approximation in an Friedmann-Robertson-Walker background. The study considered the evolution of the bubble in both a matter-dominated and a radiation-dominated Universe. According to the hoop conjecture, the collapse of false vacuum bubbles leads to the formation of primordial black holes as soon as their compactness exceeds that of a black hole during their evolution. We found that the thin-wall approximation describes well the evolution of the bubble. However, as the bubble size becomes comparable to the size of the wall, it fails to capture the final stages of the collapse during which the maximal compactness is reached. We obtained the latter from the lattice simulations, finding that the maximal compactness reached in the collapse is given by the total mass of the bubble. The requirement that the compactness exceeds that of a black hole then gives a lower-bound on the mass of the resulting black holes.

The proposed scenario thus defines a new formation mechanism that leads to a substantial production of primordial black holes, characterised by a mass function that is a relatively flat power law scaling with $M^{-1/2}$. As a result, the heavier black holes produced constitute the bulk of the resulting energy density. The upper limit on the achievable black hole masses is due to the nucleation of true vacuum bubbles within the enclosing false vacuum domains. We find that, due to this specific shape of the mass function, this scenario is too constrained to explain the observed dark matter abundance through the produced black holes. In particular, the asteroid mass window is closed by the evaporation constraints affecting the resulting extended low mass tail. We find that the collapse of false vacuum bubbles still leads to a population of black holes that can source the mergers observed by LIGO-Virgo or provide seeds for supermassive BHs. Furthermore, the produced black holes can lead to the onset of a transient matter-dominated epoch before matter-radiation equality, thereby disrupting the nucleosynthesis dynamics due to the energy injection produced in their evaporation process.

A potential feature of the proposed mechanism is that long range correlations in the spectator field fluctuations may modulate the false vacuum bubbles formation probability, which in turn can lead to a clustered spatial distribution of primordial black holes. A similar phenomenon is observed in first order phase transitions, where vacuum bubbles nucleate in clusters~\cite{Pirvu:2021roq}. Such initial clustering can strongly affect primordial black hole binary formation and, consequently, their merger rate~\cite{Raidal:2017mfl,Ballesteros:2018swv,Vaskonen:2019jpv,Young:2019gfc,DeLuca:2021hde}. 

As for the observed dark matter abundance, it is possible that modifications in the final stages of the collapse dynamics, as well as in the black holes evaporation physics, could alleviate the corresponding bound and thus allow sub-asteroid mass primordial black holes to play the role of dark matter~\cite{Raidal:2018eoo}. Alternatively, the obtained population of evaporating black holes could give rise to the observed dark matter abundance via the same evaporation process. As spectator field fluctuations can also give rise to a particle dark matter abundance~\cite{Peebles:1999fz,Markkanen:2018gcw,Tenkanen:2019aij}, the scenario can naturally accommodate a multiple-component dark sector.

The present study demonstrates that spectator fields possessing potentials with multiple minima can lead to the overproduction of primordial black holes. Therefore, the proposed scenario constitutes also a novel phenomenological probe for all standard model extensions that include new scalar degrees of freedom lighter than the inflation scale. Likewise, depending on how the standard model Higgs boson potential is stabilized, it is possible that the Higgs boson itself lead to the formation of primordial black holes through the proposed mechanism. 

As shown by the stochastic formalism, spectator field fluctuations typically generate a highly non-homogeneous distribution of domains whose structure begins to evolve after inflation ends. It is then plausible that the black holes formation be accompanied by a gravitational wave signal sourced by domain wall collisions, which may contribute to the stochastic gravitational wave background. In particular, in this paper we focused on scenarios in which the stochastic fluctuations that cause transitions between the two minima of the scalar potential are rare. Further study is needed to pinpoint the primordial black holes formation dynamics and the corresponding gravitational wave signals in scenarios where such transitions are more frequent.

\acknowledgments 
This work was supported by the Estonian Research Council grants PRG803, PRG356, by the European Regional Development Fund and the programme Mobilitas Pluss grant MOBTT86, by the European Regional Development Fund CoE program TK133 ``The Dark Side of the Universe", by the Spanish MINECO grants FPA2017-88915-P and SEV-2016-0588, the Spanish MICINN (PID2020-115845GB-I00/AEI/10.13039/501100011033), and the grant 2017-SGR-1069 from the Generalitat de Catalunya. IFAE is partially funded by the CERCA program of the Generalitat de Catalunya.

\appendix

\section{Solving the Fokker-Planck equation}
\label{app:FP}

The FP equation~\eqref{eq:FP} can be recast as a Schr\"odinger equation and studied using the equivalent eigenvalue problem~\cite{Risken}
\bea\label{eq:EIGS}
	\left[-\frac{\pd^2}{\pd \chi^2}+\mathcal{V}'(\chi)^2 - \mathcal{V}''(\chi) - \epsilon_n \right]
	\psi_n(\chi)
	=0\,, \qquad
	\epsilon_n \equiv \frac{8\pi^2v^2}{\Hinf^3} \Lambda_n\, ,
\eea
where $\mathcal{V} \equiv 4\pi^2 V/(3\Hinf^4)$, $\chi \equiv \phi/v$ and a prime denotes differentiation with respect to $\chi$. The eigenfunctions $\psi_{n}$ form an orthonormal and complete basis  $\int_{-\infty}^\infty\td\chi \, \psi_n(\chi) \psi_m(\chi) = \delta_{nm}$, $\sum_n\psi_n(\chi) \psi_n(\chi')=\delta(\chi-\chi')$. The eigenfunctions of the FP operator~\eqref{eq:FP} read $P_{n} \propto e^{-\mathcal{V}}\psi_n$ with $P_{0} \propto e^{-2\mathcal{V}}$ for non-absorbing boundary conditions. 

\begin{figure}
\centering
\includegraphics[width=0.9\textwidth]{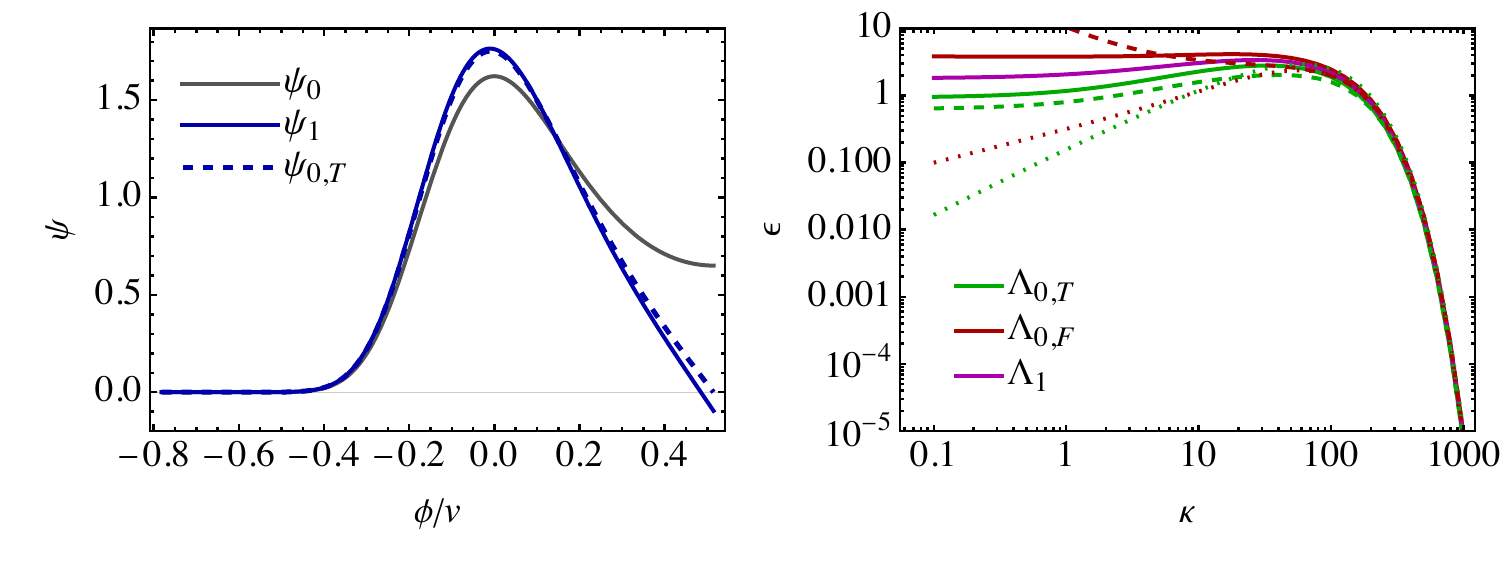}
\caption{\emph{Left panel:} The lowest eigenfunctions $\psi_n$ of Eq.~\eqref{eq:EIGS}, plotted in the true vacuum region when $\kappa = 100$. The solid lines show the eigenfunctions computed with the non-absorbing boundary conditions while the dashed line shows the eigenfunction that use an absorbing boundary condition at $\phi_b$. All functions are normalized to unity in the depicted region.  \emph{Right panel:} The leading non-vanishing eigenvalues computed with different boundary conditions. The dashed lines show the approximation in Eq.~\eqref{eq:eps_1_appr} and the dotted line the approximation in Eq.~\eqref{eq:eps_1_appr2}. Both the panels use $\Delta V/\Hinf = 0.01$.}
\label{fig:FP_sols}
\end{figure}

A high barrier between the minima implies that the lowest eigenvalue, which describes transitions between these points, is exponentially suppressed. Comparing the eigenvalues obtained with absorbing and non-absorbing boundary conditions, we see that the lowest non-vanishing eigenvalues roughly match. In fact, when $\phi_b$ is chosen to match the node of $\psi_1$ in the non-absorbing case, i.e. $\psi_1(\phi_b) = 0$, then $\psi_1$ also satisfies the absorbing boundary condition and, as a result, the lowest non-vanishing eigenvalues is substantially the same for either boundary condition. Notice also that, in case of a symmetric potential, the absorbing boundary conditions select exactly the odd eigenfunctions and eigenvalues of the corresponding non-absorbing system. Furthermore, when seeking the lowest non-vanishing eigenvalue for $\kappa \gg 1$, the absorbing and non-absorbing formulations are expected to give similar results. Nevertheless, since $\phi_b$ is chosen to coincide with the top of the potential barrier instead of the node of $\psi_1$, the eigenvalues will generally be different. A comparison between leading eigenvalues and eigenfunctions is shown in Fig.~\eqref{fig:FP_sols}.

Following Ref.~\cite{Starobinsky:1994bd} we solve for the lowest non-vanishing eigenvalue, $\epsilon_0$, by using it as small parameter in a perturbative expansion. Taking $\psi = \psi^{(0)} +  \epsilon_0 \psi^{(1)} + \ldots$ and solving $\mathcal{L}_{\rm FP} \psi = \epsilon_0 \psi$ at first order in $\epsilon_0$ with the condition $\psi(\phi_b) = 0$, we find that the lowest non-vanishing eigenvalue is approximately
\be\label{eq:eps_1_appr}
    \epsilon_{0,T} \approx \left[ \int^{\chi_b}_{0} \td \chi e^{2\mathcal{V}(\chi)} \int^{\chi_b}_{-\infty} \td \chi e^{-2\mathcal{V}(\chi)} \right]^{-1}.
\ee
Computing these integrals via the method of steepest descent gives
\be\label{eq:eps_1_appr2}
    \epsilon_{0,T} \approx \frac{2}{\pi} \sqrt{-\mathcal{V}''(0)\mathcal{V}''(\chi_b)} \,e^{-2(\mathcal{V}(\phi_b)-\mathcal{V}(0))} \,,
\ee
which holds well for large $\kappa$, as can be seen in Fig.~\eqref{fig:FP_sols}. The eigenvalues $\epsilon_{0,F}$ can be computed in a similar fashion albeit the replacements $\phi = 0 \to \phi = v$ and $\phi = -\infty \to \phi = \infty$ in the integration limits.

In terms of the rescaled eigenfunctions in Eq.~\eqref{eq:EIGS}, the transition probability~\eqref{eq:trans_Pt} can expressed as 
\be
    P(\phi, t|\phi',t') = \frac{\psi_0(\phi)}{\psi_0(\phi')} \sum_{n=0}^\infty e^{-\Lambda_n (t-t')}\, \psi_n (\phi)\psi_n (\phi')\,.
\ee
With non-absorbing boundary conditions, the equilibrium distribution $\Peq(\phi)$, obtained in the limit $t\gg t'$, is solely determined by the 0-th eigenfunction $\psi_0(\phi)$ with eigenvalue $\Lambda_0 =0$, that is, $\Peq(\phi) = \psi_0(\phi)^2 \propto e^{-2\mathcal{V}}$.

\bibliography{PBH}
\end{document}